\documentclass[final,authoryear,10pt]{elsarticle}

\usepackage{graphicx}
\usepackage{color,soul}
\usepackage{epstopdf}
\usepackage{listings,color}
\usepackage{relsize}
\usepackage{hyperref}
\usepackage{booktabs}
\usepackage{natbib} 
\sethlcolor{green}
\usepackage{amssymb}
\usepackage{amsthm}
\usepackage{amsmath}

\definecolor{verbgray}{gray}{0.9}

\lstnewenvironment{out_R}{
  \lstset{belowcaptionskip=1\baselineskip,
  backgroundcolor=\color{verbgray},
  breaklines=true,
  frame=single,
  framerule=0pt,
  xleftmargin=0pt,
  showstringspaces=false,
  basicstyle=\scriptsize\ttfamily,
  }}{}

\definecolor{shadecolor}{rgb}{.9, .9, .9}

\begin{document}
\begin{frontmatter}

\title{How does bad and good volatility spill over across petroleum markets?\tnoteref{label1}}

\author[ies,utia]{Jozef Barun\'ik\corref{cor2}} \ead{barunik@utia.cas.cz}
\author[cerge]{Ev\v{z}en Ko\v{c}enda} \ead{evzen.kocenda@cerge-ei.cz}
\author[utia,ies]{Luk\'a\v{s} V\'acha} \ead{vachal@utia.cas.cz}

\cortext[cor2]{Corresponding author} 
\address[utia]{Institute of Information Theory and Automation, Academy of Sciences of the Czech Republic, Pod Vodarenskou Vezi 4, 182 00, Prague, Czech Republic} 
\address[ies]{Institute of Economic Studies, Charles University, Opletalova 21, 110 00, Prague, Czech Republic}
\address[cerge]{CERGE-EI, Charles University and the Czech Academy of Sciences, Politickych veznu 7, 11121 Prague, Czech Republic}

\tnotetext[label1]{We benefited from valuable comments we received from Joe Brada, Ionut Florescu, and presentation participants at the Modeling High Frequency Data in Finance 5 conference (New York, 2013), the Computational and Financial Econometrics conference (London, 2013), and seminar at the Arizona State University (Tempe, 2014). This paper was written when Ev\v{z}en Ko\v{c}enda was a Fulbright Scholar at the Welch College of Business, Sacred Heart University, and its hospitality is greatly appreciated. The support of GACR grant no. 14-24129S is gratefully acknowledged. The usual disclaimer applies.} 

\begin{abstract}
We detect and quantify asymmetries in volatility spillovers using the realized semivariances of petroleum commodities: crude oil, gasoline, and heating oil. During the 1987--2014 period we document increasing spillovers from volatility among petroleum commodities that substantially change after the 2008 financial crisis. The increase in volatility spillovers correlates with the progressive financialization of the commodities. In terms of asymmetries in spillovers we show that periods of increasing crude oil prices strongly correlate with dominating spillovers due to bad volatility. Overall, bad volatility due to negative returns spills over among petroleum commodities to a much larger extent than good volatility due to positive returns. After the 2008 financial crisis the asymmetries in spillovers markedly declined in terms of total as well as directional spillovers. An analysis of directional spillovers further reveals that no commodity dominates other commodities in terms of spillover transmission in general.
\end{abstract}
\end{frontmatter}

\section{Introduction, motivation, and relevant literature}
Research on the interdependence of financial markets has led to analyzing not only returns and volatility, but their spillovers as well \citep{dimpfl2012financial}. The global financial and economic crisis, sharp fluctuations in commodity prices, and rapid financialization of petroleum commodities\footnote{The term financialization relates to investments in commodities made by investors to diversify their portfolios.} prompted a fresh surge of interest in how the dynamic links among commodities work (relevant literature is shown presently). Knowledge of spillover dynamics has important implications for investors and financial institutions in terms of portfolio construction and risk management as these spillovers and their direction may greatly affect portfolio diversification and insurance against risk \citep{gorton2005}. Further, it may be of great importance to differentiate between spillovers due to bad or good volatility \citep{bartram2012us} as the asymmetry has been proven to play an important role in many economic and financial issues related to our analysis \citep{ramos2013oil, du2011, nazlioglu2013, bermingham2011testing}. 

Petroleum-based commodities form an asset class where spillovers historically play a prominent role \citep{haigh2002crack} given importance of these commodities for the economy and economic development \citep{Hamilton1983} and the fact that shocksÕ transmissions into oil prices significantly affect the U.S. and global economy \citep{Kilian2008, hamilton1996happened, gronwald2012oil}. However, the research on volatility spillovers among petroleum commodities is rather limited and the asymmetric aspect of spillovers is not adequately explored yet. In our paper we make two key contributions. First, we use high-frequency data to extend the literature on volatility spillovers among key petroleum commodities: crude oil, heating oil, and gasoline. Second, by augmenting the current methodology of \cite{diebold2009measuring, diebold2012better}, we are able to quantify negative and positive asymmetries in spillovers, including the directions and magnitudes over time. Among other results, we rigorously show that bad volatility spills over more than good volatility across petroleum-based commodities. Such negative asymmetry coincides with periods of increasing crude oil prices. Asymmetries in spillovers considerably decline after the 2008 crisis.

Petroleum-based commodities are essential to our economies primarily from an industrial perspective.\footnote{The importance of the crude oil for our society can be documented by the 89.4 million barrels of global daily consumption in 2012 as reported by the U.S. Energy Information Administration. The corresponding figures for the largest consumption regions in millions of barrels daily are 29 for Asia, 18.5 for the US, and 14.4 for Europe; U.S. Energy Information Administration, accessed on April 24, 2014 (\url{ http://www.eia.gov/cfapps/ipdbproject/IEDIndex3.cfm?tid=5&pid=5&aid=2}).} Accordingly, crude oil prices are driven by distinct demand and supply shocks \citep{Kilian2008, hamilton2009, lombardi2011}. Further, \cite{kilian2009} shows that shifts in the price of oil are driven to different extents by aggregate or precautionary demand related to market anxieties about the availability of future oil supplies. \cite{kilian2011} support this finding by showing that energy prices do not respond instantaneously to macroeconomic news but \cite{mason2013jump} argue that the spot price of crude oil and its futures prices do contain jumps. Finally, \cite{sari2011} argue that global risk perceptions have a significantly suppressing effect on oil prices in the long run.

Besides the above forces, oil prices might also be linked to large speculations \citep{hamilton2009, caballero2008} and short run destabilizations in the oil prices may be caused by financial investors \citep{lombardi2011}. These findings are in line with petroleumÕs increasing financialization after 2001 as shown in \cite{fratzscher2013} and the expanding financialization of commodities in general \citep{mensi2013, creti2013links, dwyer2011global, vivian2012commodity}.

Due to their real economic importance and their ongoing financialization, petroleum-based commodities are naturally sensitive to economic development as well as market volatility. The recent evidence in \cite{VachaBar2012} indicates that during periods of recession there exists a much higher downside risk to a portfolio formed from oil-based energy commodities. The asymmetric risk and accompanying volatility spillovers are thus a feature one would like to measure and monitor effectively. The research related to volatility spillovers among energy commodities is surprisingly limited, though. On weekly data, \cite{haigh2002crack} analyze the effectiveness of crude oil, heating oil, and unleaded gasoline futures in reducing price volatility for an energy trader: uncertainty is reduced significantly when volatility spillovers are considered in the hedging strategy. Using daily data for the period 1986--2001, \cite{hammoudeh2003causality} analyzed the volatility spillovers of three major oil commodities (West Texas Intermediate, heating oil, and gasoline) along with the impact of different trading centers. Spillovers among various trading centers were also analyzed by \cite{awartani2012dynamic}, who investigated the dynamics of the return and volatility spillovers between oil and equities in the Gulf region. The spillover effect between the two major markets for crude oil (NYMEX and London's International Petroleum Exchange) has been studied by \cite{lin2001spillover}, who found substantial spillover effects when both markets are trading simultaneously. More recently, \cite{chang2010analyzing} have found volatility spillovers and asymmetric effects across four major oil markets: West Texas Intermediate (USA), Brent (North Sea), Dubai/Oman (Middle East), and Tapis (Asia-Pacific).

It is not surprising that different classes of petroleum commodities are affected by similar shocks given their potential substitution effect \citep{chevallier2013} or economic linkages \citep{casassus2013}. However, the spillovers might evolve differently depending on the qualitative nature of the shocks. In terms of volatility spillovers, it is of key importance to identify how negative or positive shocks transmit to other assets. Changes in the volatility of one commodity are likely to trigger reactions in other commodities. We hypothesize that such volatility spillovers might exhibit substantial asymmetries and we aim to quantify them precisely.

Much of the research studying volatility spillovers among markets have employed multivariate GARCH family models, VEC models, etc. However, these methods have interpretative limitations as, most importantly, they are not able to quantify spillovers in sufficient detail. In our analysis we build on more efficient technique. Recently, \cite{diebold2009measuring} introduced a methodology for the computation of a spillover index (DY index) based on forecast error variance decomposition from vector autoregresssions (VARs).\footnote{While the DY index has been widely adopted to analyze spillovers on financial markets, to the best of our knowledge, only one study applies the methodology to measuring volatility spillovers on the commodity markets, albeit without assessing asymmetries in spillovers. Using daily data, \cite{chevallier2013} find that volatility spillovers among commodities have been increasing in the period 1995--2012. They even show that the inclusion of commodities in a broad portfolio of assets increases total spillovers. Among the commodities, the biggest net contributors to spillovers are precious metals and energy commodities. Hence, exploring asymmetry in spillovers among key energy commodities represents an important area that has not been explored yet.} The methodology has been further improved in \cite{diebold2012better} by introduction of spillover direction and variable ordering in VARs. Another improvement of the original DY index has been introduced by \cite{klossner2014exploring}, who developed a new algorithm for fast calculation of the index along with the computation of the minimum and maximum values of the index. Finally, based on the idea of realized semivariance due to \citep{shephard2010measuring}, \cite{bkv2013} extended the information content of the DY index with the ability to capture asymmetries in spillovers that materialize due to negative and positive returns/shocks: we employ this methodology for our analysis.

Our contribution is centered on finding substantial asymmetries in volatility spillovers across petroleum commodities, but our results are much richer. 
During the 1987--2014 period we document increasing spillovers from volatility among petroleum commodities that substantially change character after the 2008 financial crisis. The increase in volatility spillovers correlates with the progressive financialization of the commodities. In terms of asymmetries in spillovers we show that periods of increasing crude oil prices strongly correlate with negative asymmetries in spillovers. Overall, bad volatility due to negative returns spills over across petroleum commodities to a much larger extent than good volatility. After the 2008 financial crisis the asymmetries in spillovers markedly declined for both total spillovers as well as directional spillovers. Analysis of directional spillovers further reveals that no commodity dominates other commodities in terms of spillover transmission.

The paper is organized as follows. In Section \ref{sec:metodology} we introduce the methodology to quantify asymmetries in volatility spillovers, namely the spillover index with realized variance and semivariance, and an intuitively appealing spillover asymmetry measure. Data of the used energy commodities are described in Section \ref{sec:data}. We display our results and inferences in Section \ref{sec:results}. Finally, we briefly conclude.

\section{Measuring asymmetries in volatility spillovers \label{sec:metodology}}

To define a measure of asymmetries in volatility spillovers, we begin with a description of the two methodological frameworks that we finally combine into a new spillover asymmetry measure.

\subsection{Realized variance and semivariance}

Consider a continuous-time stochastic process for log-prices, $p_t$ evolving over a time horizon $[0\le t \le T]$, which consists of a continuous component and a pure jump component, $p_t=\int_0^t\mu_s ds + \int_0^t\sigma_s d W_s + J_t$, where $\mu$ is a locally bounded predictable drift process and $\sigma$ is a strictly positive volatility process, and all is adapted to a common filtration $\mathcal{F}$. The quadratic variation of the log-prices $p_t$ is
\begin{equation}
[p_t,p_t] = \int_0^t\sigma_s^2 ds+\sum_{0<s\le t}(\Delta p_s)^2,
\end{equation}
where $\Delta p_s = p_s - p_{s-}$ are jumps, if present. A natural measure for quadratic variation has been formalized by \cite{andersen2001distribution} and \cite{barndorff2002econometric}, who propose to estimate it as the sum of squared returns and coined the name ``realized variance" ($RV$). Formally, let us suppose that the prices $p_0,\ldots,p_n$ are equally spaced on the interval $[0,t]$, then 
\begin{equation}
RV=\sum_{i=1}^n \left(p_i-p_{i-1}\right)^2
\end{equation}
converges in probability to $[p_t,p_t]$ with $n\rightarrow \infty$. More recently, \cite{shephard2010measuring} introduced estimators that capture the variation only due to negative or positive returns using an estimator of realized semivariance:
\begin{eqnarray}
RS^-&=& \sum_{i=1}^n \left(p_i-p_{i-1}\right)^2 I_{(p_i-p_{i-1}<0)} \\
RS^+&=& \sum_{i=1}^n \left(p_i-p_{i-1}\right)^2 I_{(p_i-p_{i-1}>0)}.
\end{eqnarray}
The realized semivariances provide a complete decomposition of the realized variance, as $RV=RS^- + RS^+$, and can serve as measures of downside and upside risk. The decomposition holds exactly for any $n$. \cite{shephard2010measuring} show the limiting behavior of the realized semivariance, which converges to $1/2\int_0^t\sigma_s^2 ds$ and the sum of the jumps due to negative and positive returns. 

\subsection{Measuring volatility spillovers \label{sec:SI}}
\cite{diebold2009measuring} introduce a volatility spillover measure based on forecast error variance decompositions from vector auto regressions (VARs). Variance decompositions record how much of the $H$-step-ahead forecast error variance of some variable $i$ is due to innovations in another variable $j$, hence the measure provides a simple intuitive way of measuring volatility spillovers. The methodology however has its limitations. First, it relies on the Cholesky-factor identification of VARs, and thus the resulting variance decompositions can be dependent on variable ordering. Second, a more crucial shortcoming of this methodology is that it allows measuring total spillovers only. Both limitations were successfully eliminated in their subsequent work, \cite{diebold2012better}, which uses a generalized vector autoregressive framework in which forecast error variance decompositions are invariant to the variable ordering, and explicitly includes the possibility to measure directional volatility spillovers. 

Third, and most important to us, \cite{diebold2009measuring,diebold2012better} use the daily or weekly range-based volatility of \cite{garman1980estimation} to compute spillovers. While range-based estimators provide an efficient way of estimating volatility, it is appealing to take advantage of the availability of high-frequency data to improve the understanding of the transmission mechanism. Due to \cite{shephard2010measuring}, we can conveniently decompose daily volatility into negative and positive semivariance providing a proxy of downside (upside) risk. Replacing the total volatility that enters the computation by the measure of downside (upside) risk will allow us to measure the spillovers from bad and good volatility and test if they are transmitted in the same magnitude. Thus, we consider $\mathbf{RV_t} = (RV_{1t},\ldots,RV_{nt})'$ to measure total volatility spillovers, and $\mathbf{RS_t^{-}} = (RS^{-}_{1t},\ldots,RS^{-}_{nt})'$ and $\mathbf{RS_t^{+}} = (RS^{+}_{1t},\ldots,RS^{+}_{nt})'$ to measure volatility spillovers due to negative and positive returns, respectively.

To measure spillovers from bad and good volatility, we use the \cite{diebold2012better} directional spillover measure, which follows directly from the variance decomposition associated with an $N$-variable vector autoregression fitted to volatility (in our case semivariances). To set the stage, consider an $N$-dimensional vector $\mathbf{RV_t} = (RV_{1t},\ldots,RV_{nt})'$ holding the realized variance of $N$ assets, which is modeled by a covariance stationary vector autoregression VAR($p$) as
\begin{equation}
\mathbf{RV_t} = \sum_{i=1}^p \mathbf{\Phi}_i \mathbf{RV}_{t-i}+ \boldsymbol{\epsilon}_t,
\end{equation}
with $\boldsymbol{\epsilon}_t\sim N(0,\mathbf{\Sigma}_{\epsilon})$ being a vector of independently and identically distributed disturbances and $\mathbf{\Phi}_i$ for $i=1,\ldots,p$ coefficient matrices. Provided that the VAR process is invertible, it has the moving average representation $\mathbf{RV}_t = \sum_{i=0}^{\infty}\mathbf{\Psi}_{i}\boldsymbol{\epsilon}_{t-i}$, where the $N\times N$ matrices holding coefficients $\mathbf{\Psi}_i$ can be obtained from the recursion $\mathbf{\Psi}_i = \sum_{j=1}^p\mathbf{\Phi}_j \mathbf{\Psi}_{i-j}$ with $\mathbf{\Psi}_0$ being the identity matrix; $\mathbf{\Psi}_0=\mathbf{I}_N$ and $\mathbf{\Psi}_i = 0$ for $i<0$. The moving average representation is key to understanding the dynamics of the system as they allow computation of variance decompositions. These in turn allow to decompose the forecast error variances of each variable in the system into parts which are attributable to various system shocks. \cite{diebold2012better} build the spillover index on the idea of assessing the fraction of the $H$-step-ahead error variance in forecasting $i$th variable that is due to shocks to $j$th variable for $j\ne i$, for each $i$. In order to obtain variance decompositions, which are invariant to variable ordering in the VAR system, \cite{diebold2012better} use the framework of generalized VAR of \cite{koop1996impulse} and \cite{pesaran1998generalized}. The framework allows for correlated shocks but accounts for them by using the observed distribution of the errors, under a normality assumption. In this way, the shocks to each variable are not orthogonalized. Hence the resulting sum of the contributions to the variance of the forecast error may not necessarily equal one.

\subsubsection{Total spillovers}
To define the spillover index, \cite{diebold2012better} consider $H$-step-ahead generalized forecast error variance decomposition matrix $\Omega$, which has following elements $\omega_{ij}^H$ for $H=1,2,\ldots$
\begin{equation}
\omega_{ij}^H=\frac{\sigma_{jj}^{-1}\sum_{h=0}^{H-1}\left( \mathbf{e}'_i \mathbf{\Psi}_h \mathbf{\Sigma}_{\epsilon}\mathbf{e}_j \right)^2}{\sum_{h=0}^{H-1}\left( \mathbf{e}'_i \mathbf{\Psi}_h \mathbf{\Sigma}_{\epsilon}\mathbf{\Psi}'_h\mathbf{e}_i \right)},
\end{equation}
 where $\mathbf{\Sigma}_{\epsilon}$ is the variance matrix for the error vector $\boldsymbol{\epsilon}_t$, $\sigma_{jj}$ is the standard deviation of the error term for the $j$th equation, $\mathbf{e}_i$ is selection vector, with one as the $i$th element and zeros otherwise, and $\mathbf{\Psi}_h$ moving average coefficients from the forecast at time $t$. The sum of the elements in each row of the variance decomposition table is not equal to one, $\sum_{j=1}^N \omega_{ij}^H\ne1$, as the shocks are not necessarily orthogonal in this framework. Hence we need to normalize each element by the row sum as $ \widetilde{\omega}_{ij}^H = \frac{\omega_{ij}^H}{\sum_{j=1}^N \omega_{ij}^H}$.
Using the contributions from the variance decomposition, \cite{diebold2012better} then define the total spillover index, which measures the contribution of spillovers from volatility shocks across variables in the system to the total forecast error variance as
\begin{equation}
\label{stot}
\mathcal{S}^H=100\times \frac{1}{N} \sum_{\substack{i,j=1\\ i\ne j}}^N\widetilde{\omega}_{ij}^H
\end{equation}
Note that by construction, $\sum_{j=1}^N \widetilde{\omega}_{ij}^H=1$ and $\sum_{i,j=1}^N \widetilde{\omega}_{ij}^H=N$, thus contributions of spillovers from volatility shocks are normalized by the total forecast error variance.

\subsubsection{Directional spillovers}

The spillover index as defined by the Eq. (\ref{stot}) helps us to understand how much of the shocks to the volatility spill over across the studied assets. The main advantage of the generalized VAR framework is, however, the possibility to identify directional spillovers using the normalized elements of the generalized variance decomposition matrix. Directional spillovers allow us to further uncover the transmission mechanism, as we can decompose the total spillovers to those coming from, or to, a particular asset in the system.

\cite{diebold2012better} propose to measure the directional spillovers received by asset $i$ from all other assets $j$ as:
\begin{equation}
\mathcal{S}_{i\leftarrow\bullet}^H=100\times \frac{1}{N} \sum_{\substack{j=1\\ i\ne j}}^N\widetilde{\omega}_{ij}^H
\end{equation}
In a similar fashion, the directional spillovers transmitted by asset $i$ to all other assets $j$ can be measured as:
\begin{equation}
\mathcal{S}_{i\rightarrow\bullet }^H=100\times \frac{1}{N} \sum_{\substack{j=1\\ i\ne j}}^N\widetilde{\omega}_{ji}^H
\end{equation}

\subsubsection{Net spillovers and net pairwise spillovers}
Directional spillovers can also be used to obtain the net volatility spillover from asset $i$ to all other assets $j$ as a simple difference between gross volatility shocks transmitted to and received from all other assets:
\begin{equation}
\label{netspills}
\mathcal{S}^H_i=\mathcal{S}_{i\rightarrow\bullet }^H-\mathcal{S}_{i\leftarrow\bullet}^H
\end{equation}
The net volatility spillover tells us how much each asset contributes to the volatility in other assets, in net terms. 

Finally, it is also interesting to define the pairwise volatility spillover between asset $i$ and $j$ as the difference between the gross shocks transmitted from asset $i$ to asset $j$ and those transmitted from $j$ to $i$:
\begin{equation}
\label{netpairwise}
\mathcal{S}^H_{ij}=100\times \frac{1}{N} \left(\widetilde{\omega}_{ji}^H-\widetilde{\omega}_{ij}^H \right)
\end{equation}

\subsection{Measuring asymmetric spillovers}

Finally, we describe how to capture and measure asymmetric volatility spillovers. Specifically, we are able to account for spillovers from volatility due to negative returns $\mathcal{S}^-$ and positive returns $\mathcal{S}^+$, as well as directional spillovers from volatility due to negative returns $\mathcal{S}_{i\leftarrow\bullet}^-$, $\mathcal{S}_{i\rightarrow\bullet}^-$, and positive returns $\mathcal{S}_{i\leftarrow\bullet}^+$, $\mathcal{S}_{i\rightarrow\bullet}^+$. Based on the previous exposition, to isolate asymmetric volatility spillovers we need to replace the vector of volatilities $\mathbf{RV_t} = (RV_{1t},\ldots,RV_{nt})'$ with the vector of negative semivariances $\mathbf{RS_t^{-}} = (RS^{-}_{1t},\ldots,RS^{-}_{nt})'$ or the vector of positive semivariances $\mathbf{RS_t^{+}} = (RS^{+}_{1t},\ldots,RS^{+}_{nt})'$. Please note that we drop the $H$ index to ease the notational burden from here on, but it remains a parameter for the estimation of spillover indices. If the contributions of $RS^-$ and $RS^+$ are equal, the spillovers are symmetric, while the differences in realized semivariance result in asymmetric spillovers. Moreover, we assume that the values of the volatility spillover indices differ over time. To capture the time-varying nature, we compute indices using a moving window.

\subsubsection{Spillover Asymmetry Measure}
In order to better quantify the extent of volatility spillovers we introduce a spillover asymmetry measure ($\mathcal{SAM}$) that is formally defined as
\begin{equation}
\label{sam}
\mathcal{SAM} = 100\times \frac{\mathcal{S}^+-\mathcal{S}^-}{1/2 \left(\mathcal{S}^++\mathcal{S}^-\right)},
\end{equation}
where $\mathcal{S}^-$ and $\mathcal{S}^+$ are volatility spillover indices due to negative and positive semivariances, $RS^-$ and $RS^+$, respectively, with $H$-step-ahead forecast at time $t$. The $\mathcal{SAM}$ defines and illustrates the extent of asymmetry in spillovers due to $RS^-$ and $RS^+$. When $\mathcal{SAM}$ takes the value of zero, spillovers coming from $RS^-$ and $RS^+$ are equal. When $\mathcal{SAM}$ is positive, spillovers coming from $RS^+$ are larger than those from $RS^-$ and the opposite is true when $\mathcal{SAM}$ is negative.

\subsubsection{Directional Spillover Asymmetry Measure}
While the spillover asymmetry measure ($\mathcal{SAM}$) defined by Eq. (\ref{sam}) measures to what extent the spillovers from volatility are asymmetric, we can decompose this measure and study the source of asymmetry among studied assets. 
We define the asymmetry measure for directional spillovers received by asset $i$ from all other assets $j$ as
\begin{equation}
\label{dsamfrom}
\mathcal{SAM}_{i\leftarrow\bullet} = 100\times \frac{\mathcal{S}_{i\leftarrow\bullet}^+-\mathcal{S}_{i\leftarrow\bullet}^-}{1/2 \left(\mathcal{S}_{i\leftarrow\bullet}^++\mathcal{S}_{i\leftarrow\bullet}^-\right)},
\end{equation}
In a similar fashion, we can measure the degree of asymmetry in directional spillovers transmitted by asset $i$ to all other assets $j$:
\begin{equation}
\label{dsamto}
\mathcal{SAM}_{i\rightarrow\bullet} = 100\times \frac{\mathcal{S}_{i\rightarrow\bullet}^+-\mathcal{S}_{i\rightarrow\bullet}^-}{1/2 \left(\mathcal{S}_{i\rightarrow\bullet}^++\mathcal{S}_{i\rightarrow\bullet}^-\right)}.
\end{equation}
$\mathcal{SAM}_{i\leftarrow\bullet}$ and $\mathcal{SAM}_{i\rightarrow\bullet}$ allow us to identify the extent to which volatility from (or to) $i$th asset spills over to (or from) other assets symmetrically. For example if bad volatility from one asset in the system transmits to other volatilities more than the positive one, $\mathcal{SAM}_{i\rightarrow\bullet}$ will be different from zero, and we expect it to be negative. This information would stay hidden in the original \cite{diebold2012better} framework.

\section{Data \label{sec:data}}

The data set consists of transaction prices for crude oil, heating oil, and gasoline traded on the New York Mercantile Exchange (NYMEX); the data were obtained from Tick Data, Inc. We use the most active rolling contracts from the pit (floor-traded) session during the main trading hours of 9:00--14:30 EST. From the raw irregularly spaced prices we extract 5-minute logarithmic returns using the last-tick method for the RV, $RS^-$, and $RS^+$ estimators. The 5-minute choice is guided by the volatility signature plot, and previous literature employing the same data. The sample period goes from September 1, 1987 through February 12, 2014. In 2006, NYMEX changed the grade of gasoline, and instead of unleaded gasoline (HU) contracts, started to trade reformulated gasoline blendstock for oxygen blending (RBOB) futures. For the gasoline data, we use unleaded gasoline until late 2006, and RBOB gasoline from 2006. We eliminate transactions executed on Saturdays and Sundays, U.S. federal holidays, December 24 to 26, and December 31 to January 2, due to the low activity on these days, which could lead to estimation bias. 

\begin{figure}
\centering
\includegraphics[scale=0.49]{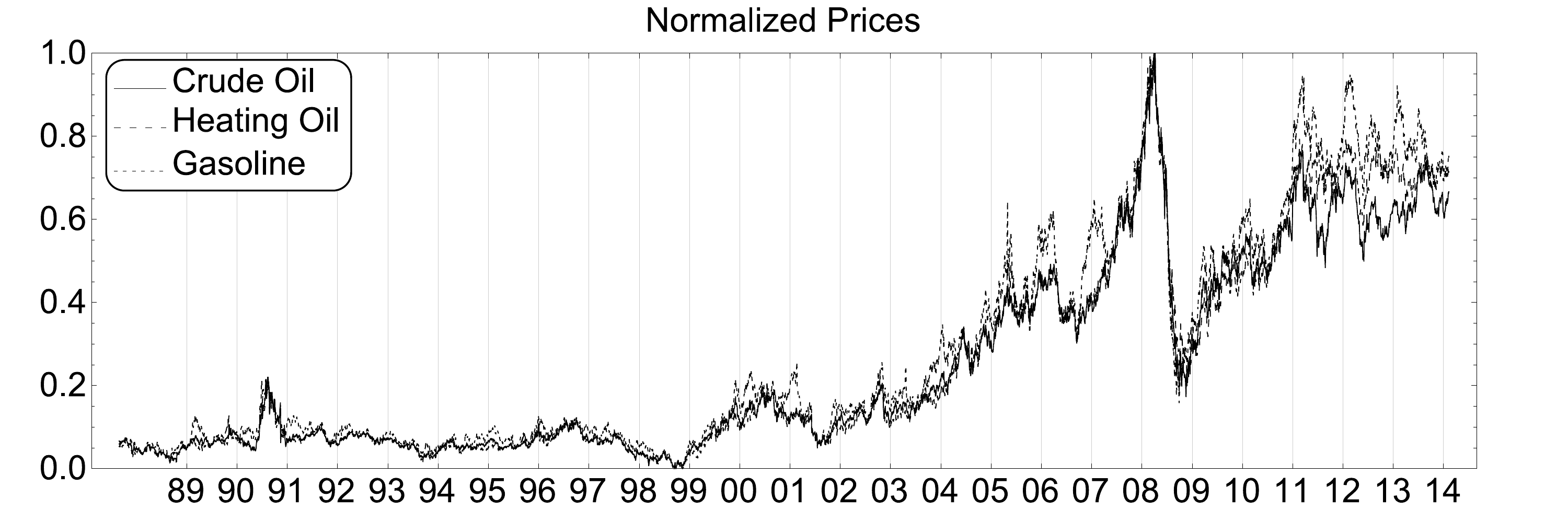}
\caption{Normalized prices of crude oil, heating oil, and gasoline over the sample period extending from September 1, 1987 through February 12, 2014.}
\label{fig:prices}
\end{figure}

Table \ref{tab:summarystats} reports the summary statistics for the estimated realized measures. The daily prices are plotted in Figure \ref{fig:prices}. 

\section{Results \label{sec:results}}
This section summarizes the results of the volatility spillover analysis of petroleum commodities. For easier orientation we divide our results into three parts. The first part shows the dynamics of spillovers and uncovers important patterns in the volatility transmission mechanism. The second part introduces asymmetries and shows the importance of understanding the differences in information transmission from bad and good volatility due to negative and positive shocks. The last part examines directional spillovers along with asymmetries.

\subsection{Extent to which uncertainty spills over petroleum markets\label{sec:results1}} 

\begin{figure}[h]
\centering
\includegraphics[scale=0.35]{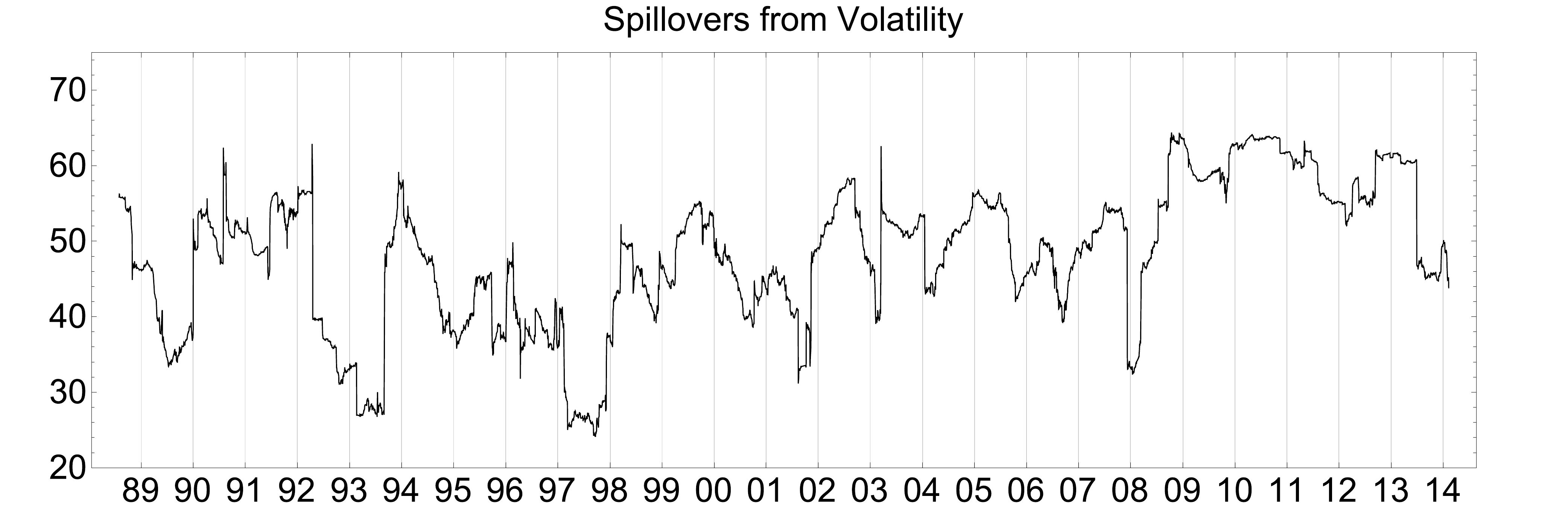} \\
\caption{Total spillover plot: spillovers from volatility. Crude oil, heating oil, and gasoline over the sample period extending from September 1, 1987 through February 12, 2014.}
\label{fig:totalspills}
\end{figure}

The total volatility spillover plot in Figure \ref{fig:totalspills} captures the dynamics of the volatility spillovers among the three commodities over the examined time period. The plot is constructed as a series of the volatility spillover estimates employing 200-day rolling windows,\footnote{The rolling window runs from point $t-199$ to point $t$. In addition to a 200-day window, we constructed the spillover index with rolling windows of 150 and 100 days to check the robustness of our results. We have also experimented with different $h$ values, and we find that the results do not materially change and are robust with respect to the window and horizon selection. These results are available upon request from the authors.} and horizon $h=10$. As the time span is 26 years, rich dynamics and important patterns emerge. 

The first intriguing observation is the strong dynamics of the spillovers between the volatility of the commodities under study. As heating oil and gasoline are products of crude oil, we would expect that any information from one of the commodities will transmit quickly to the other one.\footnote{In effect, all three petroleum commodities are tightly connected. \cite{casassus2013} explicitly define the production relationship between crude oil (input) and heating oil (output), and the complementary relationship (in production) between gasoline and heating oil. Further, heating oil is produced as a by-product when crude oil is cracked to produce gasoline. This implies another production relationship between crude oil (input) and gasoline (output). About 40 and 20 \% percent of the crude oil is refined into gasoline and heating oil, respectively.} Interestingly, Figure \ref{fig:totalspills} shows a different pattern. In total, spillovers from volatility are not so large. Average spillovers are reported in the volatility spillover Table \ref{tab:spills}, with a value of 50\%. Moreover, the time-varying spillover index exhibits a great degree of fluctuation, ranging from about 25\% to 65\% (Figure \ref{fig:totalspills}). This means that the volatility of one commodity does not necessarily excessively impact the volatility of other commodities all the time, although the petroleum commodities are fundamentally tied through the production process. An implication emerges: when trading petroleum futures the above evidence may be used to increase benefits from portfolio diversification during periods of low spillovers. We will study this interesting observation later by looking at directional spillovers, which could potentially uncover the source of the uncertainty in petroleum markets. 

Second, we are able to identify two distinct periods during which spillovers behave differently. During the first period, before 2008, the average value of spillovers is 45.4\% and fluctuates within 7\% standard deviation, while after 2008, it is 58.3\% with a considerably lower fluctuation of 5\% standard deviation. Hence 2008 is a dividing point: we can observe a structural break that is behind a change in volatility transmission mechanism.\footnote{As in \cite{zeileis2003testing}, we employ the supF testing methodology to formally identify an endogenous break in the spillovers index on September 14, 2008; the date precedes the official collapse of Lehman Brothers by one day.} The differences between the two periods are even more striking when we observe some details. The lowest levels of spillovers in 1989, 1993, 1997, and 2001 are in sharp contrast to the rest of the plotted spillovers, but at the same time the highest peaks of the spillovers before 2008 reach only the average level of the post-2008 period.

The increase of the volatility spillovers in 2002 and mainly 2008 has a parallel in rising energy commodity prices after 2002 (Figure \ref{fig:prices}). These patterns are deeply related to the financialization of the commodities during the previous years. Increased demand for commodities as portfolio investments resulted in a dramatic surge of their portfolio weights and energy commodities became important parts of index portfolios \citep{tang2012index}. According to \cite{cheng2013financialization}, investment inflows to various commodity futures indices totaled \$200 billion between 2000 and mid-2008. \cite{henderson2013new} document that between 2003 and 2011, financial commodity investments increased from \$15 to \$400 billion. Increased demand for financial commodity investments have also been advocated as a key source behind increases in energy commodity prices \citep{singleton2013investor,tang2012index,henderson2013new,hamilton2014risk} among others.

We pair the above evidence with the presented development of spillovers and claim that the process of advancing financialization of the energy commodities highly correlates with the increase of spillovers from early 2000s on, and this pattern is especially strong after the financial crisis in 2008. The increased correlation of energy and non-energy commodities through the increasing presence of index investors \citep{tang2012index} further enlarges the ground for volatility to spill over among other classes of commodities.

\subsection{Asymmetric transmission of information in petroleum markets \label{sec:results2}} 

Having a full picture of how uncertainty spills over the petroleum markets, we proceed to study possible asymmetries in the transmission mechanism. Earlier we argued that volatility spillovers might differ in their magnitude based on whether the shock originates from negative or positive returns. Based on the methodology proposed in earlier sections, we aim to compute spillovers due to bad and good volatility, and quantify to what extent petroleum markets process information asymmetrically.

\begin{figure}[h]
\centering
\includegraphics[scale=0.35]{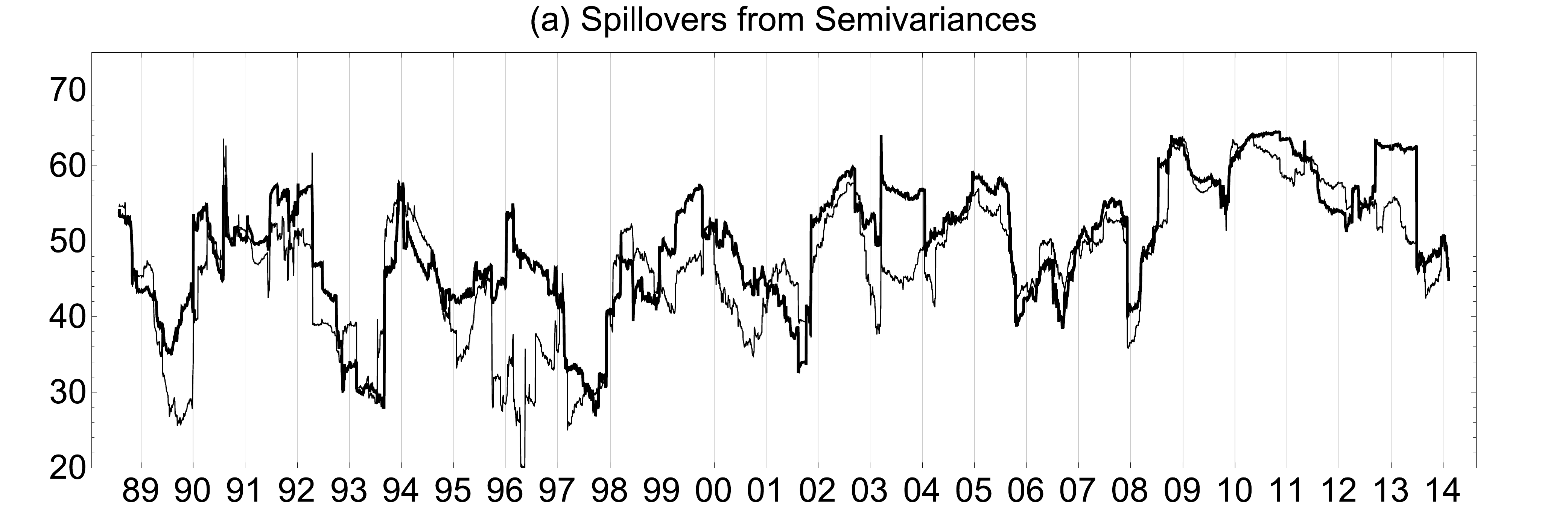} \\
\includegraphics[scale=0.35]{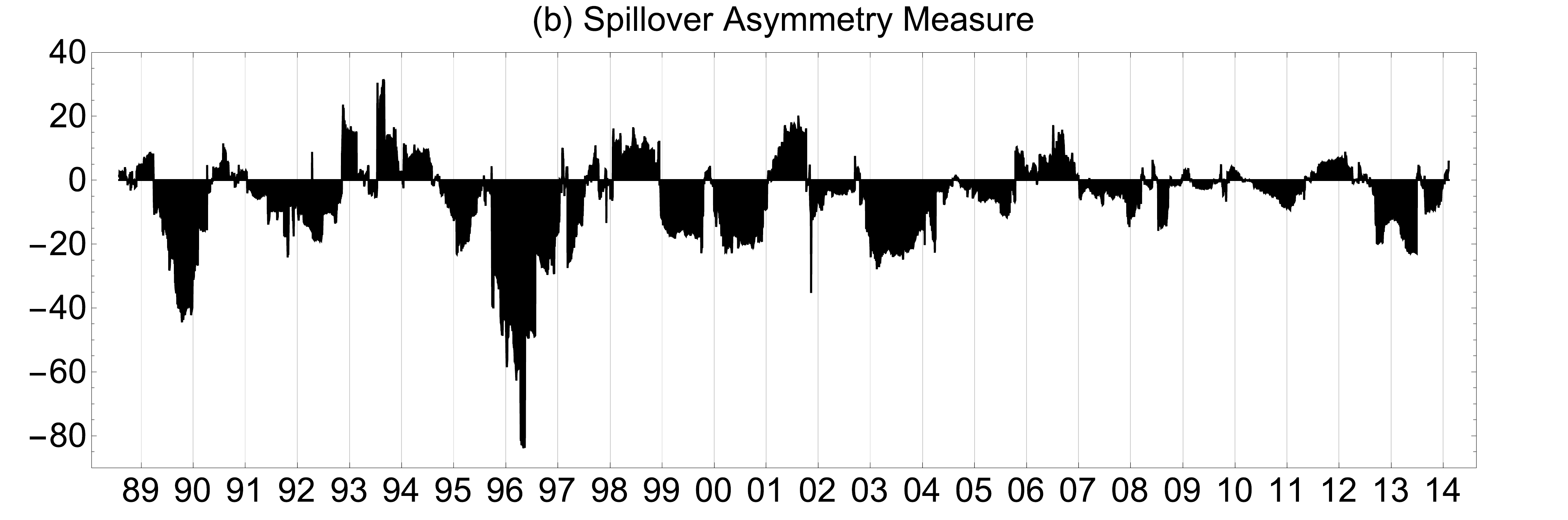}
\caption{Total spillover plot using (a) $RS^+$, and $RS^-$ (bold) semivariance for crude oil, heating oil, and gasoline over the sample period extending from September 1, 1987 through February 12, 2014. (b) Spillover Asymmetry Measure ($\mathcal{SAM}$)}
\label{fig:totalass}
\end{figure}

In panel (a) of Figure \ref{fig:totalass} we present two total spillover plots that are based on negative and positive semivariances. Hence, the plot captures patterns of total volatility spillovers that materialize due to negative and positive returns. Closer inspection of the plot reveals that both spillovers due to bad and good volatility share a common path but their developments are not identical. We can identify several periods during which spillovers due to negative and positive volatility diverge to various extents. The differences are better visible using the Spillover Asymmetry Measure ($\mathcal{SAM}$) in panel (b) of Figure \ref{fig:totalass}. 

$\mathcal{SAM}$ quantifies the differences in total volatility spillovers due to negative and positive returns and allows portraying the extent of asymmetry that is independent of the spillover levels. In case $\mathcal{SAM}$ is positive (negative) spillovers due volatility caused by positive (negative) returns dominate, while the zero $\mathcal{SAM}$ means that negative and positive information is transmitted equally between markets. A direct observation is that this neutral position of markets is very rare. The principle evidence is that asymmetry in total spillovers is overwhelmingly driven by negative returns (shocks) and the extent of asymmetricity is not only in the magnitude but also in duration. The key periods when negative returns drove volatility spillovers can be identified in 1989--1990, 1995--1997, 1999--2000, and 2003--2004; a lower impact is visible from 2008 on. Conversely, there are only a few episodes when spillovers due to positive returns are larger than their negative counterparts, plus their duration is shorter with a smaller extent. 

The first period of negative returns driving volatility spillovers in petroleum markets (1989--1990) is associated with a decrease in total spillovers. Then in 1991, a large supply shock due to the first Gulf War doubled crude oil prices in a few months; total volatility spillovers doubled as well. The most notable asymmetric effect is visible during the end of 1995 and 1996. The year 1995 was for many years the last year when the U.S. produced more oil than it imported and the U.S. dependence on foreign petroleum peaked in 2005.\footnote{ US Energy Information Administration (EIA)) (\url{http://www.eia.gov/countries/country-data.cfm?fips=US\#pet}).} Economically this is an important issue that had to be absorbed by markets and that is also in line with one of the oil-specific demand shocks peaking in 1995 and evidenced in \cite{kilian2009}. This period was followed quickly by resumed growth after the short-lived Asian Crisis. Crude oil prices rose quickly during 1999--2000 due to a large increase in consumption, and peaked before the beginning of the U.S. recession in 2001. Interestingly, the periods 1993--1994 and 2001--2002, time-wise related to these large increases in prices, were themselves marked by large decreases in prices. Positive values of $\mathcal{SAM}$ during both periods point at good volatility being transmitted to a larger extent than bad volatility. Still, the extent of positive asymmetries is much lower when compared to negative asymmetries. 

Finally, we emphasize the negative $\mathcal{SAM}$ during 2003--2004 that is associated with the second Gulf War and unrest in Venezuela. These two exogenous geopolitical events contributed to the last period during which bad news had a substantially larger influence on petroleum markets compared to good news. After 2004, oil prices increased due to increasing demand and the markets were still influenced by negative returns-based volatility spillovers more than by positive ones. However, after 2004 the magnitude of the asymmetries decisively declined. After the 2007--2008 financial crisis the absence of excessive fluctuations of volatility spillovers is even more pronounced. The low fluctuations in the $\mathcal{SAM}$ measure can be partly caused by increasing financialization. As commodities become significant parts of diversified portfolios (for example via index commodity vehicles) the risk sharing increases and the room for risk premia shrinks \citep{tang2012index}. Further, an impressive increase in the financialization of the petroleum commodities does not mean a proportional increase in the number of stocks or related assetsÕ futures. Rather, finacialization propagates via increases of portfolio sizes and the number of transactions. Increases in trading activities in particular might well induce a decline in spilloversÕ asymmetries via the price-setting mechanism on the market. As a consequence, we see higher total volatility spillovers, simultaneously with lower asymmetries between volatility spillovers induced by positive or negative shocks.

Another reason for the post-2008 symmetrical transmission of information may be that oil markets are currently in the longest period of calm volatility. After 2008, the volatility of petroleum markets decreased steadily, and currently it is at the lowest levels since the crude futures markets were established in the early 1980s. Oil prices has been rarely so stable for such a long period since the 1970s. An important factor is also the fact that OPEC suppliersÕ ability to exert market power was reduced in the 2008 turmoil and its aftermath as argued by \cite{holz2012crude}. 

Overall we may conclude that the asymmetric effects in spillovers are substantial with bad volatility due to negative shocks driving the total spillovers. 

\begin{figure}[h]
\centering
\includegraphics[scale=0.35]{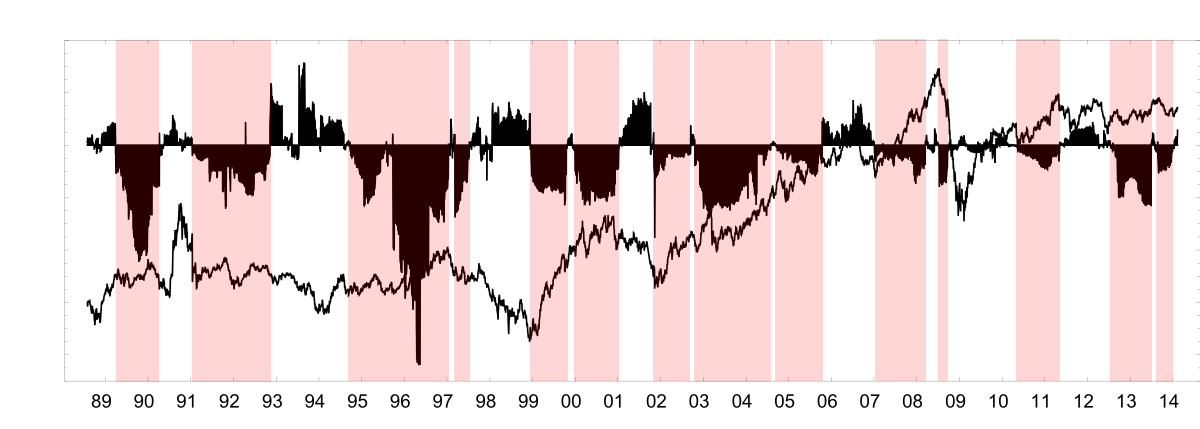} \\
\caption{Comparison plot: Spillover asymmetry measure along with crude oil prices, as a proxy for petroleum markets.}
\label{fig:comparisonplot}
\end{figure}

Finally, we add one more piece of information. While we identified clear periods of large spillover asymmetries, it may be useful to link these periods to prices, which are bound to be increasing in periods of high uncertainty. Hence, in Figure \ref{fig:comparisonplot} we present the spillover asymmetry measure along with crude oil prices, as a proxy for petroleum markets, and we highlight the periods of non-negligible negative asymmetries; the exact numbers are irrelevant in this comparison and we refrain from labeling the vertical axis of the plot. This simple link reveals a striking result: in the majority of occurrences, negative asymmetries coincide with periods when the crude oil price was increasing. Moreover, in time periods where the $\mathcal{SAM}$  is positive, the price of crude oil was declining.\footnote{Inspection of our data reveals that the effect of cold months is irrelevant.} This interesting feature in volatility spillover asymmetry may be a consequence of greater sensitivity to negative shocks in times of rising oil prices, which were usually periods of higher risk and uncertainty on the oil markets. By virtue of evidence, we show that in turbulent periods volatility spills significantly more following negative shocks (returns). 

\subsection{Directional asymmetric spillovers \label{sec:results3}}  

Earlier, we established that asymmetry in volatility spillovers among petroleum commodities is a phenomenon that does matter. We now proceed with results on asymmetries in directional spillovers; e.g. spillovers going FROM one commodity TO other commodities. The basis for the importance of the directional spillovers lies in production and complementary links among petroleum commodities. \cite{casassus2013} show that economic linkages among commodities create a source of long-term correlation between futures returns. Cross-commodity relationships and feedback-based co-movements among them form a ground for why changes in the volatility of one commodity are likely to trigger strong reactions in other commodities, and even more so in commodities of the same class.

\begin{figure}[h,t]
\centering
\includegraphics[scale=0.23]{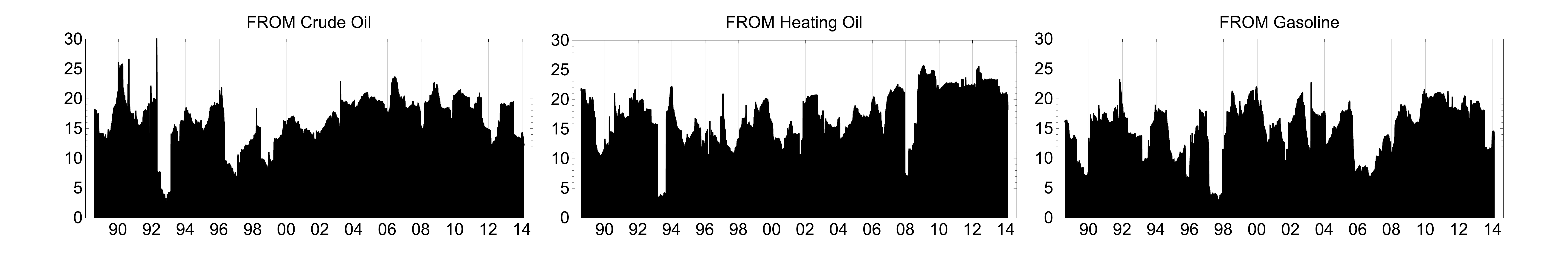}
\includegraphics[scale=0.23]{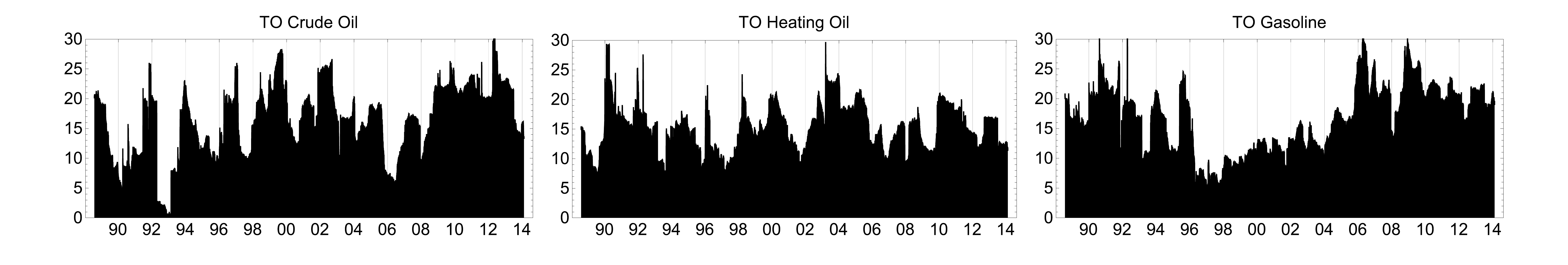}
\includegraphics[scale=0.23]{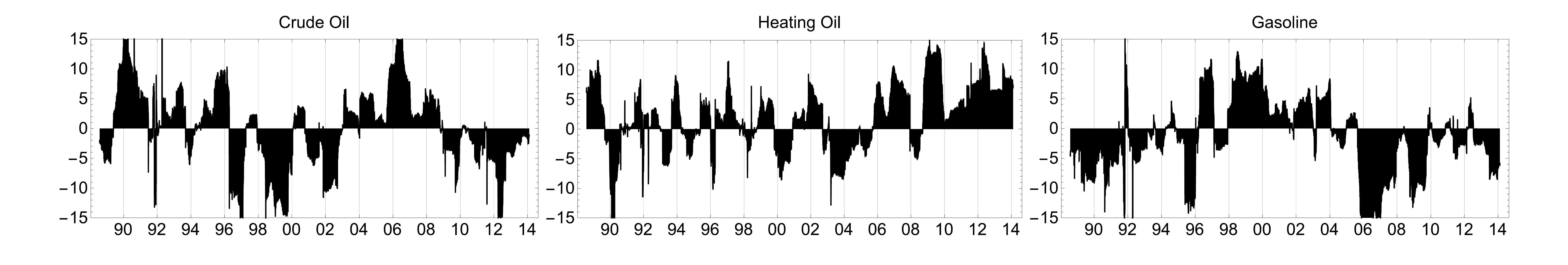}
\caption{Directional spillover plots: Directional spillovers FROM (first row), TO (second row) and Net spillovers (third row) on $RV$).}
\label{fig:dirspills}
\end{figure}

In Figure \ref{fig:dirspills}, we present directional spillovers FROM and TO a specific commodity. In the first row of the figure, we show the dynamic patterns of how a specific commodity transmits volatility to other commodities. In the second row, we demonstrate the extent of spillovers that commodities receive. In the third row we provide the net effect of the directional spillovers: a difference between ``contribution from" and ``contribution to" plotted in the first two rows. The net spillovers in the positive domain represent the position when a commodity is a spillover giver: it transmits net volatility spillovers to other commodities. The negative domain contains net spillovers that a specific commodity receives from other ones: in this case the commodity is said to be a spillover receiver. Some patterns emerge. Until 1995 crude oil was predominantly a spillover giver, then chiefly a net receiver until 2003, and again a net giver until 2008. The post-crisis period is characterized by crude oil being a spillover receiver virtually until the present. Gasoline behaves differently: it is a spillover receiver until the mid-1990s and then from 2004 on, including the 2007--2008 financial crisis period. Heating oil seems to be quite moderate in terms of transmitting and receiving net spillovers from other commodities. The net effects alternate very often and during most of the period under research net spillover values do not exceed the 5\% mark. Only after 2005 heating oil becomes a net giver and the extent of net spillovers significantly increases when compared to the previous period.

\begin{figure}[h]
\centering
\includegraphics[scale=0.23]{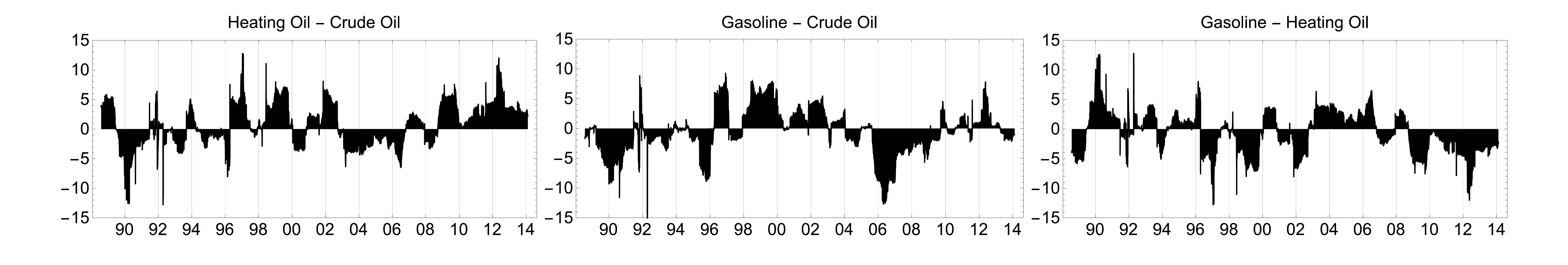}
\caption{Net pairwise spillover plots.}
\label{fig:NETpair}
\end{figure}

In Figure \ref{fig:NETpair}, we present net pairwise spillovers that show the dynamics of the net spillovers between specific pairs of commodities. The transmission of pairwise net spillovers is quite balanced in all three pairs. The key information in Figure \ref{fig:NETpair} is that no commodity dominates other commodities in terms of spillover transmission in general. The patterns of net pairwise spillovers reflect production and complementary relationships between commodities as well. 

\begin{figure}[h]
\centering
\includegraphics[scale=0.2]{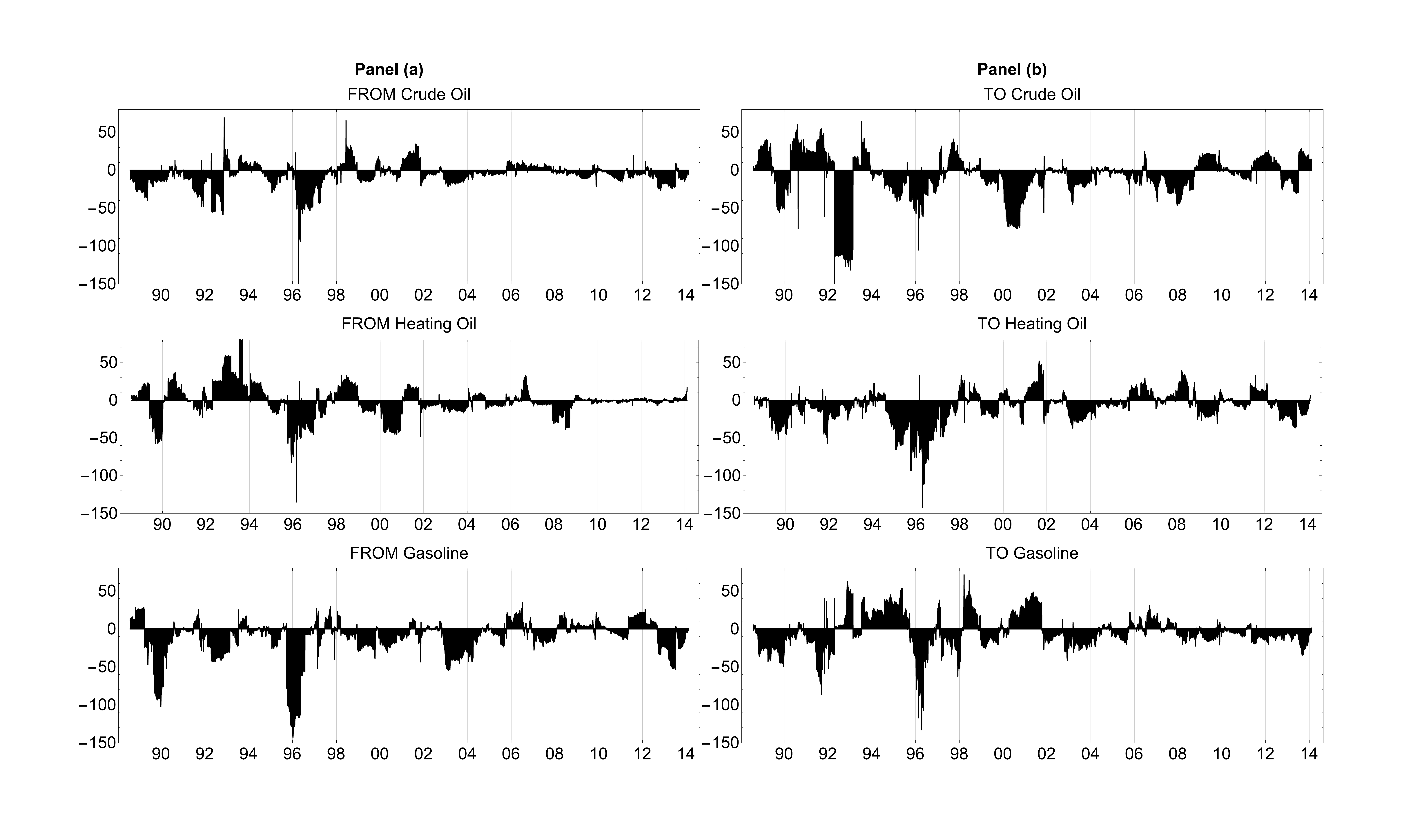}
\caption{Asymmetric directional spillover plots. Panel (a): Direction FROM ($\mathcal{SAM}_{i\rightarrow\bullet}$). Panel (b): Direction TO ($\mathcal{SAM}_{i\leftarrow\bullet}$).}
\label{fig:SAMFROMTO}
\end{figure}

Finally, the directional spillovers described above can be further decomposed to the effects that the negative and positive returns exert on volatility spillovers. In panels (a) and (b) of Figure \ref{fig:SAMFROMTO}, we present the asymmetric directional spillovers in form of plots of the directional spillover asymmetry measures ($\mathcal{SAM}_{i\leftarrow\bullet}$; panel (a) and $\mathcal{SAM}_{i\rightarrow\bullet}$; panel (b)).

The plots in Figure \ref{fig:SAMFROMTO}, panel (a), portray the dynamics of the asymmetry in spillovers FROM specific commodities outwards. There is a clear pattern of negative asymmetry that is most pronounced for the direction from crude oil and from gasoline: the positive values of $\mathcal{SAM}_{i\leftarrow\bullet}$ and $\mathcal{SAM}_{i\rightarrow\bullet}$ are small and infrequent and negative values in the case of the direction from gasoline reaches on several occasions impressive values. The dominant negative spillovers in 1992--1993 are likely associated with the steps mandated by the Clean Air Act (CAA) Amendments adopted in 1990 by the U.S. government. The specific provisions led to increases in the production of oxygenated gasoline and a number of costly adjustments  were forced on refineries and fuel distribution systems while industry profitability declined sharply and continued at low levels.\footnote{ The production of oxygenated gasoline raised chiefly in the U.S., but increases in oxygenate production capacities occurred in 1992 also in Canada, Europe, South America, and the Far East. About 31\% percent of total gasoline sales were affected during the 1992--1993 winter oxygenated gasoline season. U.S. Energy Information Administration (US EIA), 2002. Petroleum Chronology of Events 1970--2000. \url{http://www.eia.gov/pub/oil_gas/petroleum/analysis_publications/chronology/petroleumchronology2000.htm}. Accessed on November 1, 2013.} \citeauthor{guo2005oil} (\citeyear{guo2005oil};p.628) claim that ``crude oil price volatility is mainly driven by exogenous (random) events such as significant terrorist attacks and military conflicts in the Middle East". In this spirit, it is tempting to attribute large negative spillovers from crude oil and gasoline in 1996 to the disaster of the supertanker Sea Empress that caused enormous environmental damage off the coast near Wales by spilling 70,000 tons of crude oil on February 15, 1996. For the rest of the period under research the spillovers are chiefly governed by negative semivariances but their asymmetries decline after the financial crisis. This pattern is in line with our findings presented earlier.

In Figure \ref{fig:SAMFROMTO}, panel (b), we present asymmetries in spillovers TO commodities: they bring clear evidence that the directional spillovers were induced mainly by negative returns (shocks). Lengthy and often profound periods when negative returns play a key role are most visible for the direction to crude oil. Further, spillovers to heating oil exhibit a massive asymmetric effect of prolonged and deep duration for about four years (1993--1997) that can be associated with the succession of events culminating in the Asian financial crisis coupled with a decline in oil prices. Spillovers to gasoline show a relatively balanced distribution of sources divided between negative and positive returns until 2006. Afterwards, spillovers due to negative returns dominate in a mild but persistent fashion until the end of our data span. For all three commodities a common pattern of large spillovers in the negative domain is visible for example in 2003. The invasion of Iraq in 2003 prompted the interest of investors in crude oil futures markets and the ensuing extent of speculation activity heightened volatility on various markets. The Iraq War in 2003 can be behind the increased volatility on markets with crude oil \citep{Zhang2009} and is visible for other oil-based commodities. As before, asymmetries in directional spillovers decline after the financial crisis.

\section{Conclusion \label{sec:conclus}}

In this paper we study how bad and good volatility due to negative and positive returns spills over across petroleum commodities. To capture the asymmetric transmission mechanism, we combine two existing methodological approaches: the volatility spillover index of \cite{diebold2009measuring,diebold2012better} together with realized semivariances due to \cite{shephard2010measuring}. As a result we are able to detect and quantify asymmetries in volatility spillovers in high frequency data within a specific class of assets, the major petroleum commodities crude oil, gasoline, and heating oil. 

We show that, when compared to 1987--2007, volatility spillovers began to rise from the early 2000s and substantially increased after the 2008 financial crisis. At the same time the volatility spillovers became less volatile. The increase in volatility spillovers correlates with the progressive financialization of petroleum commodities after 2002. We then apply a realized semivariance approach and show a link between crude oil prices and asymmetries in spillovers: periods of increasing crude oil prices strongly correlate with negative asymmetries in spillovers. After the 2008 financial crisis the degree of (negative and positive) asymmetries markedly declines and negative and positive shocks exhibit quantitatively similar effects on volatility spillovers. Finally, an analysis of directional spillovers reveals that no commodity dominates other commodities in terms of spillover transmission in general, and asymmetries in directional spillovers decline after the financial crisis. Thus, results of directional spillovers are in line with those of total spillovers and resonate with economic relationships among the petroleum commodities. Our results also form a ground for some less-than-orthodox implications. Our findings defy a common belief that the financial crisis should prompt spillovers to be more volatile. We provide evidence of just the opposite: spillovers from price developments in 2008 and later are less volatile than before the 2007--2008 financial crisis.

{\footnotesize{
\setlength{\bibsep}{3pt}
\bibliographystyle{chicago}
\bibliography{spilloversbib}
}}

\section*{Tables and Figures}

\begin{table}[h]

\caption{Descriptive Statistics for crude oil, heating oil, and gasoline realized volatility over the sample period extending from September 1, 1987 through February 12, 2014.}
\centering
\begin{tabular}{lrrrrrr}
\toprule
& Mean & St.dev. & Skewness & Kurtosis & Minimum & Maximum\\ 
\midrule
Crude oil  &     0.3199   & 0.3465  &  4.5004  & 35.3530  &  0.1778 &   0.0056 \\
Heating oil  &     0.3042  &  0.2857  &  5.7352  & 88.6317 &   0.1780  &  0.0074 \\
Gasoline  &    0.3543   & 0.3519  &  4.7872  & 42.1653  &  0.1960  &  0.0060\\
\cmidrule{2-7}
& $\times 10^{-3}$& $\times 10^{-3}$& & & $\times 10^{-4}$ & \\ 
\bottomrule
\end{tabular}
\label{tab:summarystats}
\end{table}

\begin{table}[h]

\caption{Volatility spillover table: Rows (To), Columns (From)}
\centering
\begin{tabular}{lrrrr}
\toprule
& Crude  &   Heating Oil  & Gasoline &  \textbf{FROM}\\ 
\cmidrule{2-4}
Crude 		& 49.9025  & 21.9881  & 28.1094  & 50.0975 \\
Heating Oil 	& 25.3731  & 44.7523  & 29.8746  & 55.2477 \\
Gasoline	     & 25.1333  & 21.3211 &  53.5456  & 46.4544  \\   
& & & & \\        
\textbf{TO}	& 50.5064  & 43.3092 &  57.9839 & \textbf{TOTAL}\\
				& & 		&  	& 50.5998 \\
\bottomrule
\end{tabular}
\label{tab:spills}
\end{table}

\newpage

\end{document}